\documentclass[11pt]{article}
\newcommand{\be}{\begin{equation}}
\newcommand{\ee}{\end{equation}}

\newcommand{\ts}{\hspace{3pt}}

\usepackage{graphicx}

\begin{document}

\vspace*{.5cm}
\noindent
{\LARGE \bf  Remarks on the Compatibility\goodbreak of Opposite Arrows
of Time\footnote{\ Reply to L. Schulman's oral contribution at
the conference on the "Direction of Time", Bielefeld (2002). } }
\vskip 1.5cm
\begin{quote}
\noindent
{\bf H. D. Zeh}
\vskip 0.2cm
\noindent
Universit\"at Heidelberg\\
www.zeh-hd.de
\end{quote}

\vskip 1.4cm
{\bf Abstract:} I argue that opposite arrows of time, while being
logically possible, cannot realistically be assumed to exist during one
and the same epoch of this universe.



\section{Introduction}
If, according to the assumptions of statistical
physics, the second law is regarded as a ``fact'' rather than a
dynamical law, it could {\it conceivably} not hold at all, hold only
occasionally, or even apply in varying directions of time. Larry
Schulman demonstrated very nicely and convincingly in his oral
contribution to this conference how this may happen in principle (see
also
\cite{SchB,PRL,Naples}). The major remaining question is whether his
examples can be regarded as realistic in our universe.

In particular, we may understand from Schulman's examples how a
certain time arrow depends on ``improbable'' (low entropy)
initial or final conditions -- regardless of the direction in which we
perform our calculation. The latter (apparently trivial) remark may be
in place, since many derivations of the second law tacitly assume in a
crucial way that the calculation is used to {\it pre}dict. That
is, it is assumed to follow a ``physical'' direction of time (from an
initially given present towards an unknown future). However, precisely
this physical arrow, or the fact that only the past can be remembered
and appears ``fixed'', is a major {\it explanandum}.

While, for a given dynamical theory, we know in general precisely what
freedom of choice remains for initial {\it or} final conditions, {\it
mixed} ones (such as two-time boundary conditions) are subject to
dynamical consistency requirements -- similar to an eigenvalue
problem with given eigenvalue. This problem remains relevant even for
incomplete (for example, macroscopic) initial and final conditions.
It is usually difficult to construct an
individual solution that is in accord with both of them. In Schulman's
examples, individual solutions were mostly found by ``trial and error",
that is, by exploiting a sufficient number of solutions with given
initial conditions and selecting those which happen to fulfill the
final ones (or {\it vice versa}). However, in a realistic situation it
would be absolutely hopeless in practice ever to end up with the
required low entropy because of the exponential growth of probability
with entropy. Only an exponentially small fraction of all solutions
satisfies one or the other low entropy boundary conditions. In the
case of complete mixing, it is the square of this very small number
that measures the fraction of solutions with two-time boundary
conditions.

Being able to find solutions by trial and error
thus demonstrates already the unrealistic case. This difficulty in
{\it finding} solutions does not present any problem for their {\it
existence} on a classical continuum of states if
mixing is sufficiently complete: any set of solutions with finite
measure can be further partitioned at will, since entropy has no lower
bound in this classical situation. This conclusion is changed in
quantum theory, which would in a classical picture require the
existence of elementary phase space cells of Planck size $h^{3N}$.
The product of initial and final probabilities
characterizing the required low-entropy may then represent a phase
space volume smaller than a Planck cell -- thus indicating the absence
of any solution.

\section{Retarded and Advanced Fields}
The consistency problem in a classical setting (though without mixing)
is discussed in Schulman's ``wet carpet'' example, intended to
prove the compatibility of two interacting systems with
different, retarded and advanced, electrodynamics
\cite{PRL}. It is similar to an example studied by Wheeler and
Feynman \cite{WF48}, where a charged particle, bound to pass an
open trap door, is assumed to shut it {\it before} the particle's
arrival by means of advanced fields if it ever enters the space
behind that door. In both examples, there is but a very narrow band of
consistent solutions, in Schulman's example represented by a partly
opened window. This narrow band may exist for systems which are far
from thermodynamical mixing (cf.\ts the following sections), and they
may be consistent only if the model is considered in isolation.
In reality, macroscopic objects always interact with
their surroundings. In a causal world, this would produce ``consistent
documents'' (not only usable ones) in the thermodynamical future. In
this way, information may classically spread without limit, thus
leading to inconsistencies with an opposite arrow of other systems.
In quantum  description, classical concepts even {\it
require} the presence of irreversible decoherence, while microscopic
systems would remain in quantum superpositions of all their histories
(see Sect.\ts 6).

Philosophers are using the term ``overdetermination of the past'' to
characterize this aspect of causality \cite{Lewis, Price}. (Note that
the conventional {\it additive} physical entropy neglects such nonlocal
correlations, which would describe the consistency of documents,
for being dynamically irrelevant in the future
\cite{TD}.) In a deterministic world, one would thus have to change
{\it all} future effects in a consistent way in order to change the
past. For example, classical light would even preserve its in principle
usable memory forever in a transparent universe. In
classical electrodynamics, there is but {\it one real} Maxwell
field, while the retarded and advanced fields of certain sources are
merely auxiliary theoretical concepts. The same real field can be
viewed as a sum of incoming and retarded, or of outgoing and advanced
fields, for example (see Chap.~2 of
\cite{TD}). Retarded and advanced fields (of different sources) thus
{\it do not add}. Observing retarded fields (as our sensorium and other
registration devices evidently do) means that incoming fields related
to unspecified past sources (``noise") are negligible -- incompatible
with the presence of distinctive advanced radiation.

Problems similar to those with opposite arrows occur with closed
time-like curves (CTCs), which are known to exist {\it mathematically}
in certain solutions of Einstein's field equations of general
relativity. This existence means that {\it local} initial and final
conditions for the {\it geometry}, defined with respect to these
closed time-like curves, are identical and thus dynamically consistent.
However, CTCs are incompatible with an arrow of time, such as an
electrodynamic or thermodynamic one. Those clever science fiction
stories about time travel, which are constructed to circumvent
paradoxes, and thus seem to allow CTCs for human adventurers, simply
neglect all irreversible effects which must arise and would destroy
dynamical consistency. Since geometry and matter are dynamically
coupled, boundary conditions which lead to an arrow of time must also
protect chronology (whatever the precise dynamical model). Wet carpet
stories belong to the same category as science fiction stories: they
do not resolve the {\it unmentioned} paradoxes that would necessarily
arise from opposite arrows.

\section{Cat Maps}
Borel demonstrated long ago \cite{Borel} that microscopic states of
classically described gases are dynamically strongly
coupled even over astronomical distances. This is a time-symmetric
consequence of their extremely efficient chaotic behavior, caused by
deterministic molecular collisions. Of course, this does {\it not}
mean that macroscopic properties are similarly sensitive to small
perturbations, although their fluctuations (such as Brownian motion)
must be affected.

Macroscopic properties characterize the microscopic state of a
physical system incompletely, for example by representing a coarse
graining in phase space (or, more generally, a {\it Zwanzig projection}
-- Chap.~3 of \cite{TD}). The deterministic dynamics of initially
given coarse grains is often described by measure-preserving dynamical
maps. In contrast to deformations of extended individual objects in
space (such as Gibbs' ink drop), and even in contrast to the N
discrete points in single-particle phase space which represent a
molecular gas,  Kac's symbolic ``cats'' (areas in phase space)
\cite{Kac} represent ensembles, or sets, of {\it possible} physical
states of a given system.  Therefore, Schulman's entropy \cite{PRL} as
a function of deformed cats (``cat maps'') is  an ensemble (or
average) entropy -- not the entropy of an individual physical state.
The entropy of this ensemble is defined to depend on its distribution
in phase space, obtained after coarse graining with respect to  given
and fixed grains as a macroscopic reference system, while the entropy
of an individual state  (point in phase space) would be given solely by
the size of the specific grain that happens to contain it at a certain
time.

This is essential (and sufficient) for Schulman's argument
that the intersection of two sets representing specific initial
or final conditions is not
empty {\it if} mixing is complete. In our universe, however, some
variables participate in very strong mixing, while others (``robust''
ones, such as electromagnetic waves or atomic nuclei) may remain
stable for very long times. They are the ones that may store usable
information.

Since cat maps describe sets of states for rather simple
dynamical systems, their dynamics is far less sensitive to weak
interactions than that of individual Borel type systems. For this
reason, two systems described by cat maps with opposite arrows of time
may even be consistent for mild interactions
  \cite{PRL}.
However, these cat maps do {\it not} form a realistic model
appropriate to discuss thermodynamical arrows in our universe.

\section{(Anti-)Causality}
In order to define causality without presuming a direction of time,
one has to refer to the internal structure of the
evolving dynamical states. The above-mentioned
over-determination of the past (in other words, the existence and
consistency of multiple documents) is a typical example. Another one
is given by the concentric waves emitted from a local source. In our
world, both are {\it empirically} (not logically) related to a time
direction.

While one may expect that all such internal structures can be
shown to evolve in time from appropriate initial conditions, they are
too complex to be investigated in terms of Schulman's simple models.
For example, retarded (concentrically outgoing) waves exist in the
presence of sources  precisely when incoming fields are negligible.
This can be the case ``because'' of an initial condition for the
fields, or because of the presence of thermodynamic absorbers
\cite{TD}.

Instead of these specific structures in the states,
Schulman studied the ``effect'' (in both directions of time) of
``perturbations'' defined by small terms changing the
Hamiltonian at a certain time only
\cite{Naples}. This effect is not easily
defined in a time-symmetric way, since an ``unperturbed solution''
defined on one temporal side of the perturbation would be exclusively
changed on the other one (no matter which is the future or past). If
the unperturbed solution obeyed a two-time boundary condition, the
perturbed one would in general violate it on this ``other'' temporal
side.  In contrast to the above mentioned internal structures, our
conventional concept of perturbations is based on the
time direction used in the definition of external {\it operations}.

Therefore, in a first step, Schulman considered {\it sets} of
solutions again. The set of {\it all} solutions obeying the
``left'' boundary condition (in time) remains unchanged on the left
of the perturbation, while the opposite statement is true on the
right. However, individual solutions found in the
intersection of these two sets
(consisting of those ones which fulfill both boundary conditions) in
the case of a perturbation are generically different from the
unperturbed ones on {\it both} sides of the perturbation. Now, if
mixing is essentially complete on the right of the perturbation (that
is, for a sufficiently distant right boundary), the right boundary
condition does not affect the solutions which form the intersection
(considered as a set) on the left. This means that {\it mean values of
macroscopic variables} in the set of all solutions that are compatible
with both boundary conditions may only differ on the right (a
consequence regarded as {\it causality} by Schulman)
\cite{Naples}.  This is true, in particular, for the {\it mean}
entropy (if the latter is defined as a function of macroscopic, that
is coarse-grained, variables). Individual solutions can {\it not} be
compared in this way, since there is no individual relation between
them. In the case of complete mixing on the right, there is even a
small but non-empty subset of solutions of the original two-time
boundary value problem which keep obeying the right boundary condition
without being changed on the left. However, using them for the
argument would mean that only very specific solutions {\it can} be
perturbed in this specific sense.

In a second approach, Schulman studies
the ``effect'' of macroscopic perturbations on {\it individual}
solutions of an integrable system. This system is defined as consisting
of a finite number of independent oscillators with different
frequencies. Although solutions which fulfill both boundary conditions
can be found with and without an appropriate perturbation, they are
again not individually related. Therefore, the
causal interpretation of the perturbation remains obscure. (For {\it
closed} deterministic systems, any perturbation would itself have to be
determined from microscopic boundary conditions, and the consistency
problem becomes even more restrictive than for just two boundary
conditions.)

\begin{figure}[h]
\centering\includegraphics[width=\textwidth]{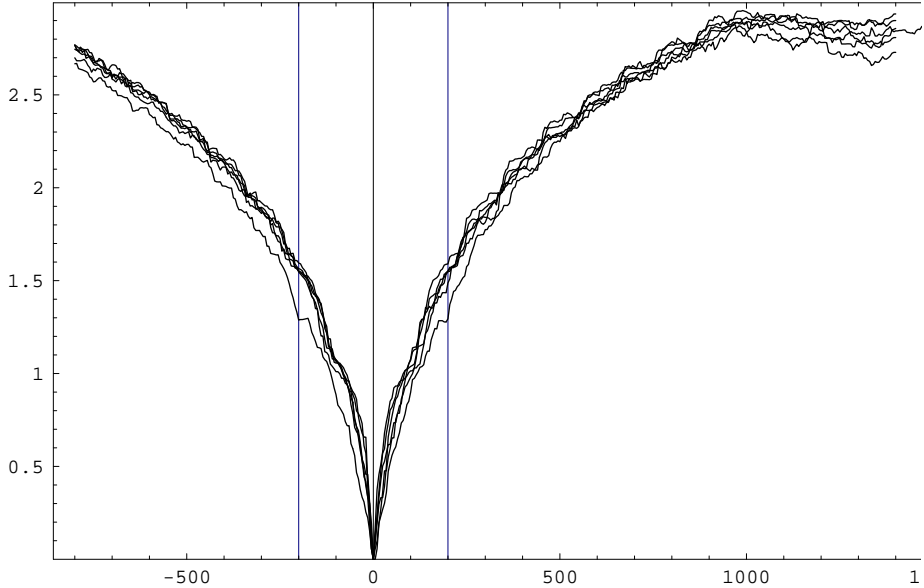}
\caption{ Four random two-time boundary solutions (forming a
narrow bundle in the diagram) are compared with two special ones,
selected by trial and error for their slightly lower entropy values at
$t_0 = 200$ or
$t_0 = -200$. Values for $t<0$ are identical with those at
$t_f-t=200.000-t$, although the final condition is actually irrelevant
in the range shown. Entropy scattering around
$t = 1300$ is accidental. (See
Appendix of \cite{TD} for details of the model and an elementary {\it
Mathematica} program for your convenience.)}
\end{figure}

Nonetheless, I was pleased to discover that Schulman's model is
formally identical with a model of particles freely moving on a
periodic interval (a ``ring'') that I had used in an appendix of
\cite{TD} for much larger numbers of constituents than used by him
(such that finding two-time boundary solutions by trial and error would
be hopeless). Particle positions on the ring have merely to be
re-interpreted as oscillator amplitudes in order to arrive at
Schulman's picture. I used this opportunity to
search by trial and error among analytically constructed two-time
boundary solutions for those ones which happen to possess {\it
slightly} lower entropy than the mean at some given ``perturbation
time''
$t_0$ (see Fig.~1). Unfortunately, the results do {\it not} confirm
Schulman's claim that these solutions are ``affected'' by the
perturbation only in the direction away from the relevant low entropy
boundary (that is, towards the ``physical future'') \cite{Naples}.
Evidently, this concept of causality, defined by means of
perturbations, is insufficient. The very concept of  a
``perturbation'' seems to be ill-defined for two-time boundary
conditions.

\begin{figure}[h]
\centering\includegraphics[width=.8\textwidth]{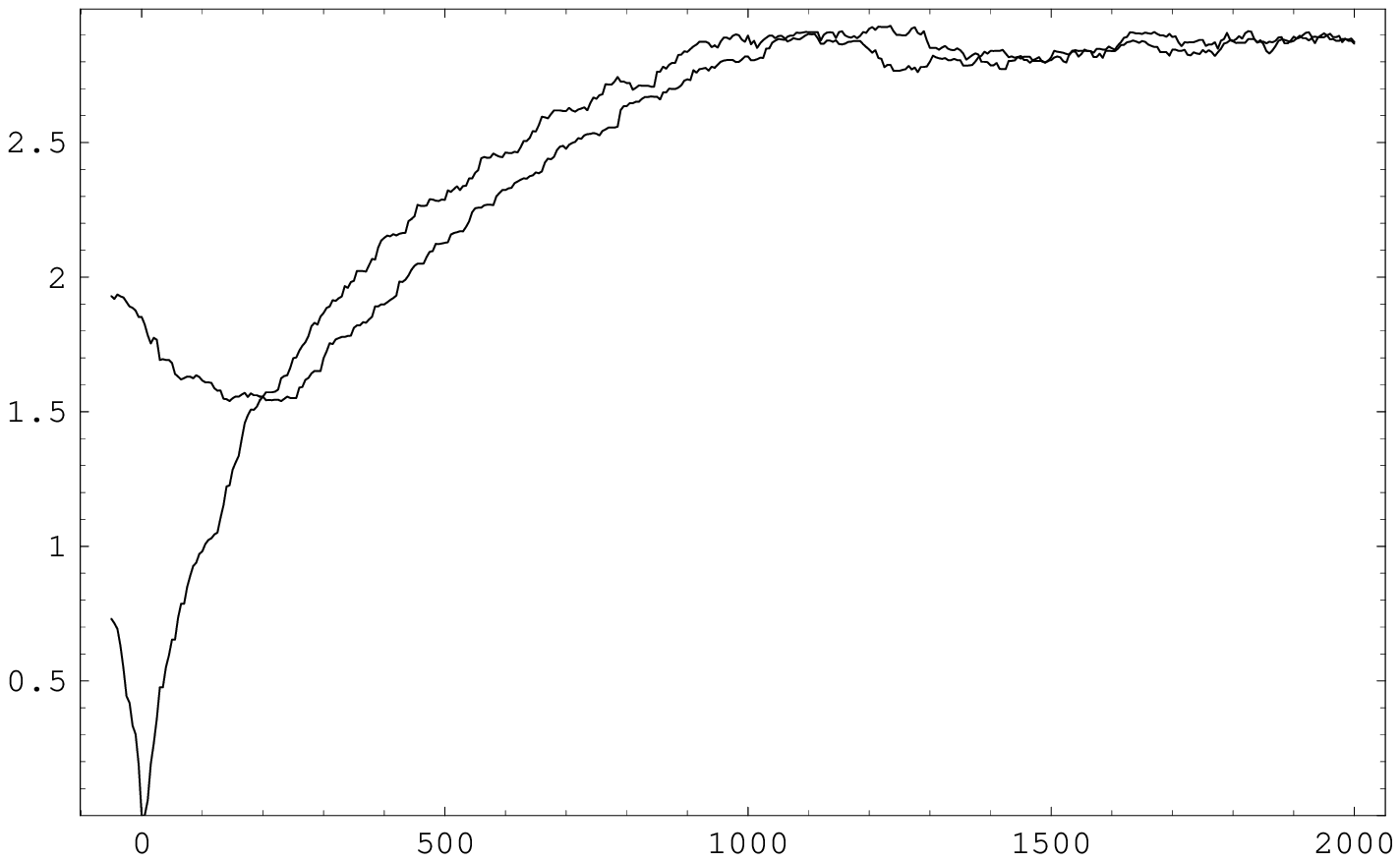}
\caption{Time-symmetric ``effect'' on a solution ``caused'' by a
perturbation of the {\it microscopic} state at time $t_0=200$,
defined by an accidental entropy minimum at this time. The perturbed
solution drastically violates {\it both} boundary conditions that were
valid for the unperturbed solution.}
\end{figure}

As another example, I calculated the effect on the solution
in both directions of time that results from a {\it microscopic}
perturbation {\it of the state} (in this case simply defined by an
interchange of velocities between particles at some time
$t_0$). Both boundary conditions are then violated by the new
solution arising from this perturbed state, used as an ``initial''
condition. The results (shown in Fig.~2) are now most dramatic towards
the former ``past'', demonstrating the relevance of fine-grained
information for correctly calculating ``backwards in time''. Deviations
from the original two-time boundary solution close to
$t_0$ also on the right are due to the fact that the
coarse graining assumed in this model does not define a very good
master equation (as discussed in \cite{TD}).

\section{Cosmology and Gravitation}
It appears evident from the above discussion that opposite arrows of
time would in general require almost complete thermalization between
initial and final conditions, which is hard to accomplish even in a
cosmological setting. In its present stage, this universe is very far
from equilibrium. A reversal of the thermodynamical arrow of time
together with that of cosmic expansion, as
suggested by Gold \cite{Gold}, would therefore require a total life
time of the universe vastly larger than its present age.  Weakly
interacting fields may never thermodynamically ``mix'' with the rest
of matter
\cite{Dyson}.

In particular, Davies and Twamley demonstrated \cite{DT} that an
expanding and recontracting universe would remain essentially
transparent to electromagnetic waves between the two radiation eras.
This means that advanced radiation resulting from all stars which will
exist during the recontraction of our universe would be present now,
apparently unrelated to any individual future sources because of their
distance, but red- {\it or} blue-shifted, depending on the size of the
universe at the time of (time-reversed) emission. According to Craig
\cite{Craig}, this radiation would show up as a non-Planckian
high-frequency tail of the cosmic background radiation resulting from
the past radiation era (where it would be absorbed in time-reverse
description). This leads to the consistency problems described in
Sect.\ts 2.

While neutrinos from the future would presumably remain unobserved,
gravity, despite its weakness, dominates the entropy capacity
of this world, and leads to consequences which are the most difficult
ones to reverse. Black holes are expected to harbor event
horizons which would not be able ever to disappear in classical
relativity, while in quantum field theory they are predicted to
disappear into Hawking radiation in the distant future in an
irreversible manner. However, contraction of gravitating objects,
including the formation of black holes, requires that higher
multipoles are {\it radiated away}. This radiation arrow is the basis
of the ``no hair theorem'', which would characterize the asymptotic
final states of black holes in an asymptotically flat and time-directed
universe. Because of the diverging time dilation close to a horizon,
any coherent advanced radiation (with the future black hole as its
retarded cause) would be able to arrive in time to prevent the
formation of an horizon
\cite{KZ}. This solution of the ``information loss paradox'' may save a
deterministic universe (without leading to inhomogeneous
singularities).

While the required initial and final conditions are not obviously
consistent in this classical scenario, this problem is relaxed in
quantum cosmology.

\section{Quantum Aspects}
Realistic models of physical systems require quantum theory to be
taken into account. Since quantum entropy is calculated from
the density matrix (that may result from a wave function by means of
generalized ``coarse graining"), its time dependence has in principle
to include a collapse of the wave function during measurements or other
``measurement-like'' situations (such as fluctuations or phase
transitions). If the collapse represents a fundamental irreversible
process, it defines an arrow of time that is {\it never} reversed.
Only a universal Schr\"odinger equation (leading to an Everett
interpretation) could be time (or CPT) symmetric. A reversal of the
time arrow would then require decoherence to be replaced by
recoherence: advanced Everett branches must combine with our world
branch in order to produce local coherence. Although being far more
complex than a classical model (since relying on those infamous ``many
worlds'') this would still allow us to conceive of a two-time boundary
condition for a {\it global} wave function (see Sect.~4.6 of
\cite{TD}). Note that Boltzmann's statistical correlations (defined
only for ensembles) now become quantum correlations (or entanglement,
defined for {\it individual} quantum states).
For example, re-expanding black holes, mentioned in the previous
section, would in an essential way require (and possibly be
facilitated by) recoherence.

In {\it quantum gravity} (or any other ``reparametrization-invariant''
theory), the Schr\"odinger equation is reduced to the
Wheeler-DeWitt equation, $H \Psi = 0 $, which does not explicitly
depend on time at all. However, because of its hyperbolic
form, this equation defines an ``intrinsic initial value problem'' with
respect to the expansion parameter $a$. In a classical
(time-dependent) picture, the initial and final states would have to
be identified in order to define  {\it one} boundary condition,
while the {\it formal} final condition
(with respect to $a$) for recontracting universes is reduced to the
usual normalizability of the wave function for $a \to \infty$. Big bang
and big crunch (distinguished by means of a WKB time, for example)
could not even conceivably be different as (complete) quantum states
(Chap.~6 of
\cite{TD}), while forever expanding universes might be said to define
an arrow of time that never changes direction during a WKB history.
All arrows thus seem to be strongly entangled.

\end{document}